\documentclass[a4paper]{article}
\usepackage[margin=25mm]{geometry}
\usepackage{graphicx}
\usepackage[cmex10]{amsmath}
\usepackage{amssymb}
\usepackage[hidelinks]{hyperref}
\usepackage{booktabs}
\usepackage{multirow}
\usepackage[numbers,sort&compress]{natbib}
\usepackage{eurosym}

\providecommand{\keywords}[1]
{
  \small	
  \textbf{\textit{Keywords---}} #1
}

\title{Security-Constrained AC/DC Grid Optimal Power Flow Considering Asymmetrical HVDC Grid Operation using Sparse Tableau Formulation}
\author{Oscar Damanik, Giacomo Bastianel, Dirk Van Hertem, Hakan Ergun \\
        \small Department of Electrical Engineering, KU Leuven, 3001 Heverlee \\
        \small Etch, EnergyVille, 3600 Genk \\
        \small Belgium
}
\date{}

\begin{document}

\maketitle

\begin{abstract}
This paper presents a security-constrained optimal power flow (SCOPF) model for HVDC grids that optimizes the asymmetrical operation of bipolar converter stations, i.e., different current injections of the positive and negative converter poles, to minimize operational costs under post-contingency conditions caused by single converter pole outages.
The optimization model allows the selection of the number of converter stations that operate asymmetrically.
The results indicate that increasing the number of asymmetrical stations lowers operational costs.
The analysis also provides insight into the sensitivity of these costs to the level of asymmetrical operation.
However, increased asymmetrical operation leads to higher DC neutral voltage offsets that can rise to undesired levels.
Imposing limits on these offsets can, in turn, increase operational costs.
To mitigate these effects, a neutral line switching (NLS) strategy is proposed for the post-contingency state.
\end{abstract}

\section{Introduction}

\subsection{Background}

Europe has set ambitious targets for offshore wind development as part of its Green Deal and climate neutrality objectives for 2050~\cite{EU_Green_Deal_2019}.
According to ENTSO-E's Offshore Network Development Plans (ONDP)~\cite{entso-eOffshoreNetworkDevelopment2024}, coordinated offshore grid expansion across European sea basins will be essential to integrate the targeted large-scale offshore wind capacities.
Moreover, the ONDP highlights that achieving these targets depends on the development of integrated and meshed offshore HVDC networks.
Previous studies on offshore grids ~\cite{windeuropeOffshoreGridsNext2023, PROMOTioNProgressMeshed} also emphasize that meshed offshore HVDC grids are key to improving reliability and cost efficiency of the future European power system by facilitating large-scale wind integration and cross-border energy exchange.

The voltage source converter (VSC) technology is the preferred choice for these meshed HVDC networks, primarily due to its capability for independent active and reactive power control and its suitability for long-distance subsea transmission \cite{PROMOTioNProgressMeshed}.
Throughout this paper, the term “HVDC grid” refers specifically to the VSC-based HVDC system.

Developing such coordinated meshed HVDC grids requires advanced planning and a thorough understanding of their operation to ensure both economic and technical performances.
However, several challenges related to the reliable and flexible operation of such grids remain to be explored, which this study attempts to advance.
In particular, this study investigates the operational challenge where asymmetrical operation may be required after a contingency.
To address this, we introduce an optimization model that determines the optimal asymmetrical operation in meshed HVDC grids.

\subsection{Problem statement and relevant work}

Meshed HVDC grids can be realized in several configurations, depending on the type and arrangement of the converter stations they comprise~\cite{letermeOverviewGroundingConfiguration2014}.
Typically, the converter station has either monopolar or bipolar configurations.
It is also possible to combine them, for instance, tapping one or more monopolar configurations to a bipolar HVDC grid~\cite{letermeOverviewGroundingConfiguration2014,serrano-silleroHVDCGridsHeterogeneous2019}.

The bipolar configuration is often preferred for large-scale applications due to its higher reliability, as each pole can continue operating independently following a fault on the other if a return path is available~\cite{hertemHVDCGridsOffshore2016}.
The return path or neutral connection can be established either through the ground or a dedicated metallic return (DMR), which is usually preferred for offshore applications.
During normal operation, both poles transfer equal but opposite currents, resulting in zero current through the neutral connection. This is called symmetrical or balanced operation, which is enforced by the bipole control implemented at the station level~\cite{IECHVDC2023-1,INTEROPERA_D2.1}.

However, a single-pole outage in a bipolar HVDC grid leads to unequal currents between the positive and negative poles when the healthy pole remains in service, hence asymmetrical operation~\cite{serrano-silleroHVDCGridsHeterogeneous2019}, which causes currents to be injected into the neutral connection.
This asymmetrical operation is synonymous with the unbalanced operation definition used in~\cite{jatHybridACDC2024}.
Therefore, the terms asymmetrical/symmetrical and unbalanced/balanced are used interchangeably in this paper.

To continue the operation of the fault-affected converter station after a converter pole contingency, at least another station has to operate asymmetrically.
If the meshed HVDC grid includes multiple bipolar converter stations, we can select which stations operate asymmetrically, along with the setpoints of their respective converter poles.
Additionally, some stations can be restricted to symmetrical operation if required.

In light of this condition, the proposed model includes discrete decision variables that determine converter stations operating asymmetrically following any single converter pole outage.
This allows the optimization model to identify the most cost-effective combination of converters to operate under post-contingency conditions.

In the post-contingency state, nonzero DC neutral voltages can be induced across the neutral points due to the current flowing through the neutral connection. 
This condition has been investigated in ~\cite{jatHybridACDC2024}, where a metric called ``current unbalance factor'' (CUF) is proposed to measure the level of unbalanced current flowing in the HVDC grid.
It is shown that an increase in this current also raises the voltages of the neutral points ~\cite{jatHybridACDC2024}.
If not properly managed, these voltages can adversely affect the HVDC system and its subsystems, as the neutral points are typically used as a voltage reference.
This phenomenon can be referred to as DC neutral voltage distortion~\cite{IECHVDC2023-1}.

Additionally, to maintain the voltages at acceptable levels, the converter station setpoints may need to be adjusted, as the neutral currents in the grid primarily depend on the coordination of these setpoints.
However, such action can reduce operational flexibility and increase operational costs, as converters must operate under tighter constraints to satisfy these conditions.

Therefore, we propose topological actions on the neutral connection of the HVDC grid as a non-costly alternative to mitigate such a situation. This topological action consists of disconnecting one or more DMRs in the grid. In analogy to the optimal transmission switching (OTS) as applied in AC grids, we refer to it as DC neutral line switching (NLS).

To model the proposed topological action, we adopt a security-constrained optimal power flow (SCOPF) framework based on the Sparse Tableau Formulation (STF) presented in~\cite{parkSparseTableauFormulation2019}.
This approach is well-suited for representing topological actions because the STF approach efficiently handles the node-breaker network representation, which is essential for evaluating different network configurations and switching actions~\cite{babaeinejadsarookolaeeGridOperationEnhancement2025}.
In this study, we extend the STF approach, classically used in AC grids, to DC grid components.
Furthermore, we incorporate both asymmetrical operation selection and topological actions as station-level post-contingency corrective actions in the model.

\subsection{Main contributions and scope of this work}

The main contributions of this work can be summarized as follows:
\begin{itemize}
\item \textbf{STF for HVDC grids:} A novel DC-grid counterpart of the STF is developed and implemented to represent asymmetrical operation and topological actions in the proposed optimization model.
\item \textbf{Converter station selection modeling:} The proposed optimization model introduces decision variables that determine which bipolar converter stations operate asymmetrically or symmetrically following a converter pole outage.
\item \textbf{Neutral line switching (NLS):} The model introduces a novel HVDC topological strategy referred to as NLS to mitigate the cost impact associated with limiting DC neutral voltage offsets.
\item \textbf{Integrated analysis:} The results and analysis demonstrate a way to quantify the operational and economic considerations associated with asymmetrical converter station operation.
\end{itemize}

Although the focus of this work is on bipolar HVDC grid configurations, we still provide modeling for different converter configurations that are included in the test case.
This allows for a more realistic representation of large-scale interconnected systems by incorporating AC network coupling and extended HVDC connections.
The analysis focuses on steady-state post-contingency operation following permanent converter pole outages. Several case studies are performed to quantify the trade-offs between operational cost, asymmetrical converter station operation, and DC neutral voltage offset.

The remainder of this paper is organized as follows. Section~\ref{sec:methodology} presents the formulation of the proposed optimization model, including the representation of asymmetrical converter station operation and the implementation of the NLS.
Section~\ref{sec:case_study} starts with the introduction of the case study setup.
Then, it continues with the analysis of the results, highlighting the impacts of asymmetrical operation and topological actions on system performance and operating costs.
Finally, Section~\ref{sec:conclusion} concludes the paper and outlines potential directions for future research.

\section{Methodology}
\label{sec:methodology}

\subsection{STF-based formulation for HVDC networks}
The power flow problem is formulated using the STF to represent the node-breaker topology of the network~\cite{parkSparseTableauFormulation2019}.
The STF for HVDC network elements can be expressed as:
\begin{equation}
    \begin{bmatrix}
        \mathbf{0} & \mathbf{0} & A^{dc} \\
        -(A^{dc})^T & \mathbf{I} & \mathbf{0} \\
        \mathbf{0} & F_{u}^{dc} & F_{i}^{dc}
    \end{bmatrix}
    \begin{bmatrix}
        U^{dc} \\
        u^{dc} \\
        i^{dc}
    \end{bmatrix}
    =
    \begin{bmatrix}
        I^{dc} \\
        0 \\
        0
    \end{bmatrix}
    \label{eq:dc_stf}
\end{equation}
where $F_{u}^{dc}$ and $F_{i}^{dc}$ are the network element matrices, $A^{dc}$ is the node-to-element incidence matrix, $\mathbf{I}$ and $\mathbf{0}$ are the appropriate identity and zero matrices.
The network element matrices describe the two-port network representation of power system components.
For the components in an HVDC grid, the network element matrices can be obtained from their linear element equations in the form of
\begin{equation}
    F_u^{dc}
    \begin{bmatrix}
        u_i^{dc} \\
        u_j^{dc} 
    \end{bmatrix}
    +
    F_i^{dc}
    \begin{bmatrix}
        i_i^{dc} \\
        i_j^{dc} 
    \end{bmatrix}
    =
    \begin{bmatrix}
        0 \\
        0
    \end{bmatrix}.
    \label{eq:le_generic}
\end{equation}

\begin{figure}[htbp!]
    \centering
    \includegraphics[width=0.3\columnwidth]{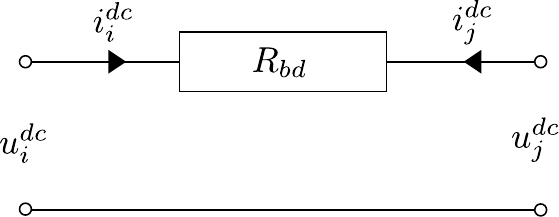}
    \caption{Two-port network representation of DC lines.}
    \label{fig:brdc}
\end{figure}

The series resistance model depicted in Fig.~\ref{fig:brdc} is used to represent DC transmission lines.
The corresponding linear element equation is given in~\eqref{eq:brdc}, from which the matrices $F_u^{dc}$ and $F_i^{dc}$ are obtained.
The operational status of each DC line is then described by the parameter $\gamma$, defined as
\begin{align*}
\gamma = 
\begin{cases}
1, & \text{if the line is in service} \\
0, & \text{if the line is out of service}.
\end{cases}
\end{align*}
For transmission switching or topology optimization problems, $\gamma$ is modeled as a binary decision variable rather than a fixed parameter.
\begin{equation}
    \underbrace{\begin{bmatrix}
        \gamma & -\gamma \\
        0 & 0 
    \end{bmatrix}}_{\textstyle F_u^{dc}}
    \begin{bmatrix}
        u_i^{dc} \\
        u_j^{dc} 
    \end{bmatrix}
    +
    \underbrace{
    \begin{bmatrix}
        1-\gamma & R_{bd} \\
        \gamma & 1 
    \end{bmatrix}}_{\textstyle F_i^{dc}}
    \begin{bmatrix}
        i_i^{dc}\\
        i_j^{dc} 
    \end{bmatrix}
    =
    \begin{bmatrix}
        0 \\
        0
    \end{bmatrix}
    \label{eq:brdc}
\end{equation}

\begin{figure}[htbp!]
    \centering
    \includegraphics[width=0.35\columnwidth]{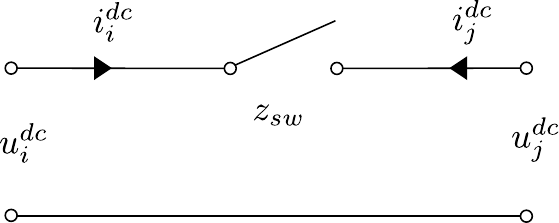}
    \caption{Two-port network representation of DC switches.}
    \label{fig:swdc}
\end{figure}

The model of DC switches is shown in Fig.~\ref{fig:swdc}.
Hence, the linear element equation is expressed as~\eqref{eq:swdc}.
Similar to the DC lines, we introduce a binary variable (or parameter) denoted by $z_{sw}$ to represent the status of the switch, where
\begin{align*}
z_{sw} = 
\begin{cases}
1, & \text{if the switch is closed} \\
0, & \text{if the switch is open}.
\end{cases}
\end{align*}
The modeling of the DC switch is included for completeness and generality, even though it is not implemented in the case study as decision variables, as we represent the DC line connection status using $\gamma$.
Nevertheless, the switching model can be employed for the same purpose by placing switches at both ends of the DC line.
Additionally, this can particularly be useful for HVDC substation switching applications, e.g., busbar splitting application presented in~\cite{bastianelOptimalTransmissionSwitching2024,morsyCorrectiveSoftBusbar2025a,bastianelOptimalTransmissionSwitching2025}.
\begin{equation}
    \underbrace{\begin{bmatrix}
        z_{sw} & -z_{sw} \\
        0 & 0 
    \end{bmatrix}}_{\textstyle F_u^{dc}}
    \begin{bmatrix}
        u_i^{dc} \\
        u_j^{dc} 
    \end{bmatrix}
    +
    \underbrace{
    \begin{bmatrix}
        1-z_{sw} & 0 \\
        z_{sw} & 1 
    \end{bmatrix}}_{\textstyle F_i^{dc}}
    \begin{bmatrix}
        i_i^{dc}\\
        i_j^{dc} 
    \end{bmatrix}
    =
    \begin{bmatrix}
        0 \\
        0
    \end{bmatrix}
    \label{eq:swdc}
\end{equation}

\subsection{Converter station modeling}
Next, we establish the modeling of the converter station.
Each converter consists of two terminals on the DC-side, denoted by 1 and 2, following the convention in~\cite{IECHVDC2023-1}.
Each terminal is connected to a DC bus, either the pole or neutral buses.
For convenience, we use the negative value of the rated voltage as the base voltage for the negative pole, while the positive pole and the neutral connection are assigned the positive value of the rated voltage.

\subsubsection{Bipolar configuration with a DMR conductor}
\begin{figure}[htbp!]
    \centering
    \includegraphics[width=0.32\columnwidth]{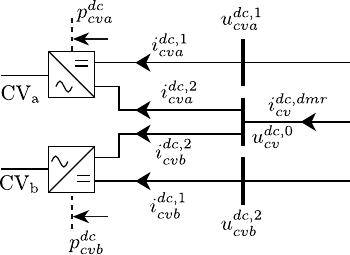}
    \caption{Schematic of a bipolar converter station with DMR.}
    \label{fig:bipolar_dmr}
\end{figure}

A bipolar converter station consists of two converters, one for each pole, as depicted in Fig.~\ref{fig:bipolar_dmr}.
Suppose $\text{CV}_\text{a}$ and $\text{CV}_\text{b}$ are the positive and negative pole converters, respectively.
The following equations provide the relationships between the currents:
\begin{equation}
    i_{cva}^{dc,1}=-i_{cva}^{dc,2}
    \label{eq:bipolar current_1}
\end{equation}
\begin{equation}
    i_{cvb}^{dc,1}=i_{cvb}^{dc,2}
\end{equation}
\begin{equation}
    i_{cv}^{dc,dmr}=i_{cva}^{dc,2}+i_{cvb}^{dc,2}
\end{equation}
Since the negative pole and neutral connections use opposite-sign voltage bases, the return current $i_{cvb}^{dc,2}$ of converter $\mathrm{CV_b}$ has the same sign as its pole current $i_{cvb}^{dc,1}$.

During asymmetrical operation in the post-contingency state, symmetrical bipole control can be enforced on a converter station by adding the following constraint:
\begin{equation*}
    i_{cva}^{dc,2}+i_{cvb}^{dc,2}=0
    \label{eq:bipolar_balance}
\end{equation*}
Therefore, the converter station will not inject any current to the neutral conductor, hence $i_{cv}^{dc,dmr}=0$. 
The DC-side power balance constraints for both converters are then calculated as
\begin{equation}
p_{cva}^{dc}=u_{cva,1}^{dc}i_{cva,1}^{dc}+u_{cv,0}^{dc}i_{cva,2}^{dc}
\end{equation}
\begin{equation}
p_{cvb}^{dc}=u_{cvb,1}^{dc}i_{cvb,1}^{dc}+u_{cv,0}^{dc}i_{cvb,2}^{dc}
\end{equation}

If one bipolar converter station is operated asymmetrically, at least another asymmetrical converter station is required to deal with the neutral current.
Suppose there is a restriction on the number of converter stations operating in asymmetrical or unbalanced mode, limited to $N_{unb}$ out of all of the total $N$ in bipolar converter stations in the grid.
The number of converters operating in symmetrical or balanced mode is then given by
$$N_{b}=N-N_{unb}$$

We model the decision of whether a converter station operates in symmetrical mode by introducing a binary decision variable $\beta_{cs}$ into the bipole control constraint for converter station $cs$.
Accordingly, the relationship is expressed as
\begin{equation}
    \beta_{cs}(i_{cva}^{dc,2}+i_{cvb}^{dc,2})=0
\end{equation}
The number of symmetrical bipolar converter stations  is constrained by
\begin{equation}
    \sum_{cs\in \mathcal{CS}}\beta_{cs}=N_{b}
    \label{eq:constraint_beta}
\end{equation}
If desired, the equality sign in~\eqref{eq:constraint_beta} can be replaced with an inequality ($\geq$) to specify $N_b$ as a minimum number of balanced converter stations.

\subsubsection{Symmetrical monopolar configuration}
\begin{figure}[htbp!]
    \centering
    \includegraphics[width=0.3\columnwidth]{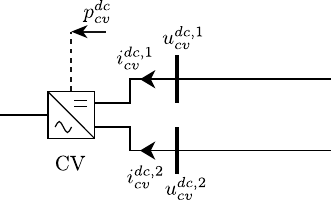}
    \caption{Schematic of a symmetrical monopolar converter station.}
    \label{fig:symm_mono_hvdc}
\end{figure}

The schematic of a monopolar converter station is shown in Fig.~\ref{fig:symm_mono_hvdc}.
Due to the symmetrical operation, we impose $i_{cv,1}^{dc}=i_{cv,2}^{dc}$.
Similar to the bipolar converter station, the signs of both currents are identical due to the opposite choice of the pole base voltages.
The DC-side power balance constraint is then generally defined as
\begin{equation}
p_{cv}^{dc}=u_{cv,1}^{dc}i_{cv,1}^{dc}+u_{cv,2}^{dc}i_{cv,2}^{dc}
\label{eq:monopole_pdc}
\end{equation}
If $u_{cv,1}=u_{cv,2}$, then $p_{cv}^{dc}=2u_{cv,1}i_{cv,1}$.

\subsubsection{DC-DC converter (symmetrical monopole)}
\begin{figure}[htbp!]
    \centering
    \includegraphics[width=0.3\columnwidth]{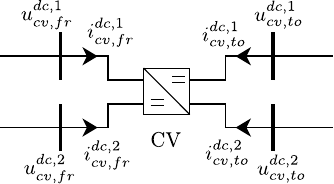}
    \caption{Schematic of a DC-DC converter station.}
    \label{fig:dcdc_conv}
\end{figure}

We include the model of a symmetrical DC–DC converter as part of the test system used in the case study.
The converter configuration is shown in Fig.~\ref{fig:dcdc_conv}.
Since the converter is symmetrical, the DC currents at both terminals are equal, i.e.,
\begin{equation}
    i_{cv,1}^{dc} = i_{cv,2}^{dc}
    \label{eq:dcdc_current}
\end{equation}
An idealized converter model is assumed; therefore, the DC-DC converter losses are neglected.
The DC-side power balance constraint also follows~\eqref{eq:monopole_pdc}.
We denote the arcs of DC-DC converters with $\mathcal{A}^{cd}$.

\subsection{SCOPF model}

The SCOPF model is described by (\ref{eq:objective_scopf})-(\ref{eq:neutral_binary}).
The objective function~\eqref{eq:objective_scopf} minimizes the operational cost, which comprises the total generation cost and the upward and downward reserve costs for the anticipated post-contingency conditions.
We denote the pre-contingency state by the index $\kappa=0$ and the post-contingency states as $\kappa\in\mathcal{K}=\{1,2,\dots,N_\kappa\}$, respectively, where each $\kappa$ corresponds to an N-1 contingency scenario.
To formulate the problem as an OPF, the reserve terms in the objective function and their associated constraints are removed, along with all variables and constraints related to the contingency scenarios.
\allowdisplaybreaks 
\begin{subequations}
    \begin{align}
        \min \begin{split}
           & \sum_{g\in\mathcal{G}}\left( C_{g}p_{g,0} + C_{g}^{up}r_{g}^{up} + C_{g}^{down}r_{g}^{down} \right) 
           \label{eq:objective_scopf}
        \end{split}   \\
        \mathrm{s.t.} \quad &\nonumber \text{hybrid STF AC power flow constraints \cite{parkSparseTableauFormulation2019}}\\&\nonumber \text{relevant AC/DC grid constraints from \cite{ergunOptimalPowerFlow2019}}\\
        &\nonumber \text{converter station constraints~\eqref{eq:bipolar current_1}-\eqref{eq:dcdc_current}}\\ 
        & F_{u,\kappa}^{dc} u_\kappa + F_{i,\kappa}^{dc} i_\kappa = 0, \quad \forall \kappa \in \{0\} \cup \mathcal{K}  \label{eq:le_dc}\\
        & I_{m,\kappa}^{dc}-A^{dc}i_\kappa = 0, \quad \forall m \in \mathcal{M}_d,\forall \kappa \in \{0\} \cup \mathcal{K} \label{eq:nte_i}\\
        & u_{\kappa} - (A^{dc})^TU_{m,\kappa} = 0, \quad \forall m \in \mathcal{M}_d,\forall \kappa \in \{0\} \cup \mathcal{K} \label{eq:nte_u} \\
        \begin{split} I_{m,\kappa}^{dc} + \sum_{cv\in\mathcal{CV}^1_m}i_{cv,\kappa}^{dc,1}+\sum_{cv\in\mathcal{CV}^2_m}i_{cv,\kappa}^{dc,2} \\      +\sum_{cv,m,n\in\mathcal{A}^{cd}}\left(i_{cv,m,n,\kappa}^{dc,1}+i_{cv,m,n,\kappa}^{dc,2}\right)= 0, \quad \\\forall m \in \mathcal{M}_d, \forall \kappa \in \{0\} \cup \mathcal{K} \label{eq:i_balance_dc} \end{split}\\
        & p_{g,\kappa}-p_{g,0} \le r_{g}^{up}  , \quad \forall g \in \mathcal{G},\forall \kappa \in \mathcal{K}\label{eq:res_pg_up} \\
        & p_{g,0}-p_{g,\kappa} \le r_{g}^{down}  , \quad \forall g \in \mathcal{G},\forall \kappa \in \mathcal{K}\label{eq:res_pg_down}\\
        & 0 \le r_{g}^{up} \le \overline{P}_g-p_{g,0} , \quad \forall g \in \mathcal{G}\label{eq:res_up_b}  \\
        & 0 \le r_{g}^{down} \le p_{g,0} , \quad \forall g \in \mathcal{G}\label{eq:res_down_b} \\
        &  \beta_{cs,\kappa} \in \{0,1\},\quad \forall cs \in \mathcal{CS},\forall \kappa \in \mathcal{K} \label{eq:bipolar_status} \\
        &  \gamma_{bd,\kappa} \in \{0,1\},\quad \forall bd \in \mathcal{BD}^{neutral},\forall \kappa \in \mathcal{K} \label{eq:neutral_binary}
    \end{align}
\end{subequations}
\allowdisplaybreaks 
The standard equality and inequality constraints for the AC/DC grids can be found in \cite{jatHybridACDC2024,parkSparseTableauFormulation2019,ergunOptimalPowerFlow2019}.
Equation~\eqref{eq:le_dc} is the linear element equations of the HVDC grid.
Equations~\eqref{eq:nte_i} and~\eqref{eq:nte_u} are the node-to-element incidence matrices of the network, describing the connectivity between the power system elements and the DC nodes in $\mathcal{M}_d$.

Equation~\eqref{eq:i_balance_dc} is the DC nodal current balance considering DC lines and converter current injections.
This approach is chosen to represent the nodal current at the neutral points appropriately \cite{jatHybridACDC2024}.
The power balance constraints of the HVDC grid are provided by the converter station constraints~\eqref{eq:bipolar current_1}-\eqref{eq:dcdc_current}.

The generator reserve constraints~\eqref{eq:res_pg_up}-\eqref{eq:res_pg_down} couple the pre- and post-contingency states.
The constraints~\eqref{eq:res_up_b} and \eqref{eq:res_down_b} limit the generator reserves, where $\overline{P}_g$ is the generator's maximum active power.
Lastly, \eqref{eq:bipolar_status} and \eqref{eq:neutral_binary} are the integrality constraints of the binary decision variables for the bipolar asymmetrical operation $\beta_{cs}$ and the DC neutral line connection status $\gamma_{bd}$, respectively.

\begin{figure*}[htbp!]
    \centering
    \includegraphics[width=1\textwidth]{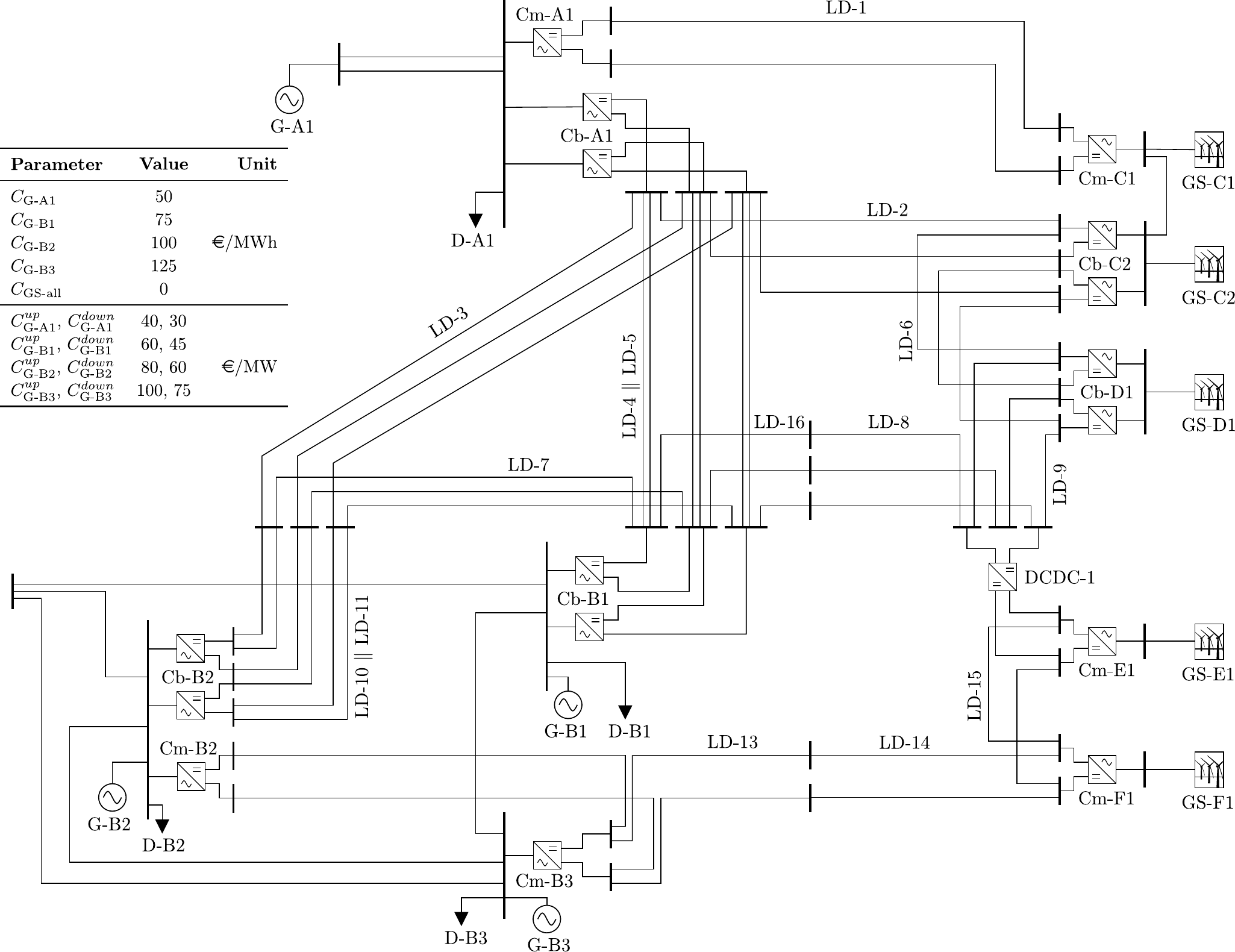}
    \caption{Modified CIGRE B4 DC grid test system in its DC multi-conductor representation. The positive, neutral (if applicable), and negative terminals or conductors always go from left to right (horizontally) or top to bottom (vertically). The cost parameters are provided on the left side for compactness.} 
    \label{fig:cigreb4}
\end{figure*}

\section{Case study}
\label{sec:case_study}
\subsection{Setup}
\label{sec:setup}
The test case is shown in Fig.\ref{fig:cigreb4}, which is adopted from \cite{vranaCIGREB4DC2013,damanikOptimalConverterControl2024} and modified by replacing the offshore load with a 500 MW offshore wind farm and using only one DC-DC converter.
In this case study, the generation costs of the offshore wind farms are set to zero.
As the problem is a mixed-integer nonlinear program (MINLP), the optimization problem is solved using Juniper, with Ipopt (3.14.19, with the linear solver ma27) employed as the NLP solver and Gurobi (11.0.3) as the MIP solver.

The proposed optimization model is validated using the selected HVDC grid test system.
The analysis aims to illustrate how the model captures the operational impacts of post-contingency asymmetrical converter station operation and NLS.
An optimal power flow (OPF) analysis is first conducted to provide intuition and insight into how the operational costs vary with respect to the different numbers of converter stations operating asymmetrically.
This is followed by the cost analysis using the SCOPF model.
Finally, the full asymmetrical operation is examined, and the NLS strategy is demonstrated as a mitigation against DC neutral voltage offsets.

\subsection{Post-contingency operations following a converter pole outage in a bipolar HVDC Grid}
In this section, we perform an OPF analysis that allows the selection of asymmetrical converter stations under a restricted number of symmetrical converters $N_b$ following a converter pole outage Cb-A1.
Therefore, we include the binary decision variable $\beta$ to ensure we obtain the optimal set of asymmetrical converter stations.
The objective is to gain insight into how the total operating cost varies with respect to $N_b$. 

\begin{figure}[htbp!]
    \centering
    \includegraphics[width=0.6\linewidth]{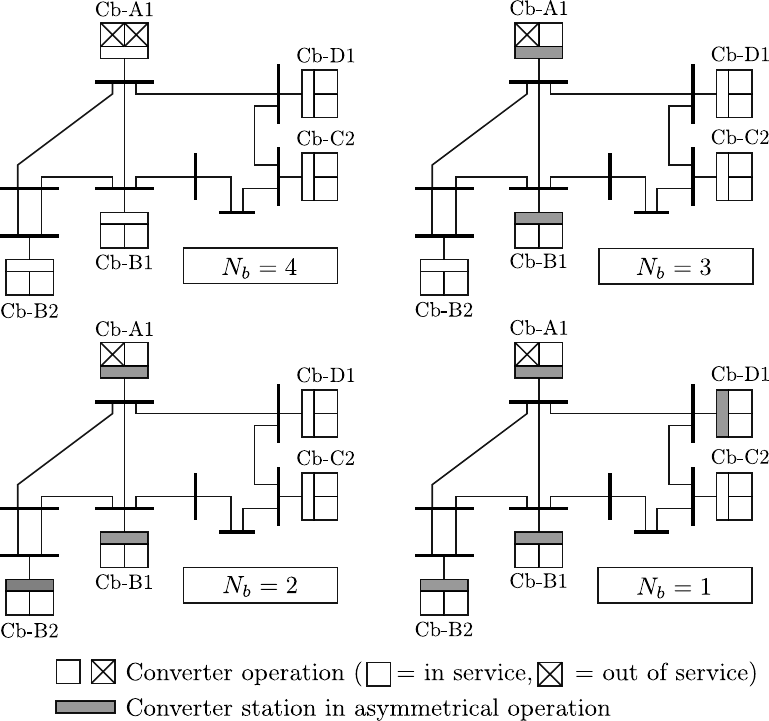}
    \caption{The decisions on symmetrical and asymmetrical converter station operation after a pole outage at Cb-A1, for different numbers of symmetrical converter stations ($N_b$) imposed. The case $N_b=0$ is omitted for brevity, since the converter selection is trivial.}
    \label{fig:unbalanced_illustrations}
\end{figure}

The compact visualization in Fig.\ref{fig:unbalanced_illustrations} illustrates the converter stations selected for asymmetrical operation for each $N_b$.
We observe that the selected converter stations prioritize the onshore terminals.
As asymmetrical operation imposes limitations on the power that can be transferred by the converters, it is less desirable to select the offshore converter stations Cb-C2 and Cb-D1 for asymmetrical operation, as they are connected to offshore wind farms with lower generation costs.

\begin{figure}[htbp!]
    \centering
    \includegraphics[width=0.6\linewidth]{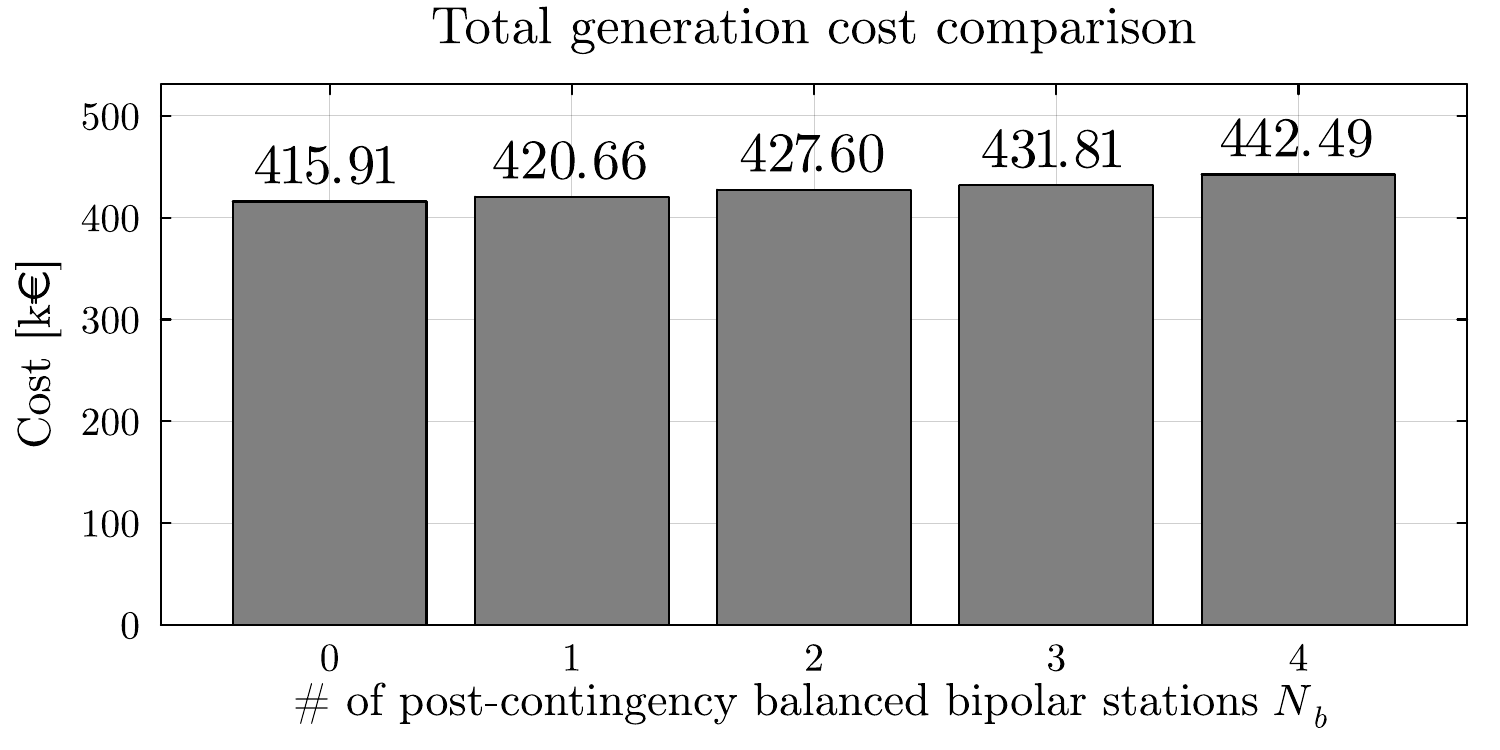}
    \caption{Post-contingency cost comparison under different numbers of balanced converter station $N_{b}$.}
    \label{fig:cost_compare_OPF}
    \vspace{-4mm}
\end{figure}

The operational costs obtained from the OPF results are then compared across these configurations as depicted in Fig.~\ref{fig:cost_compare_OPF}.
As expected, allowing more converter stations to operate asymmetrically provides greater flexibility in power redistribution and reduces the total operational cost.

\subsection{Integrating asymmetrical converter operation selection as a post-contin-gency corrective action in the SCOPF model}

We consider the N-1 contingency scenarios in the SCOPF model, where each scenario represents a pole outage in one of the bipolar converter stations.
For this, we exclude the case where $N_b=4$ because it is considered trivial.

\begin{figure}[htbp!]
    \centering
    \includegraphics[width=0.6\linewidth]{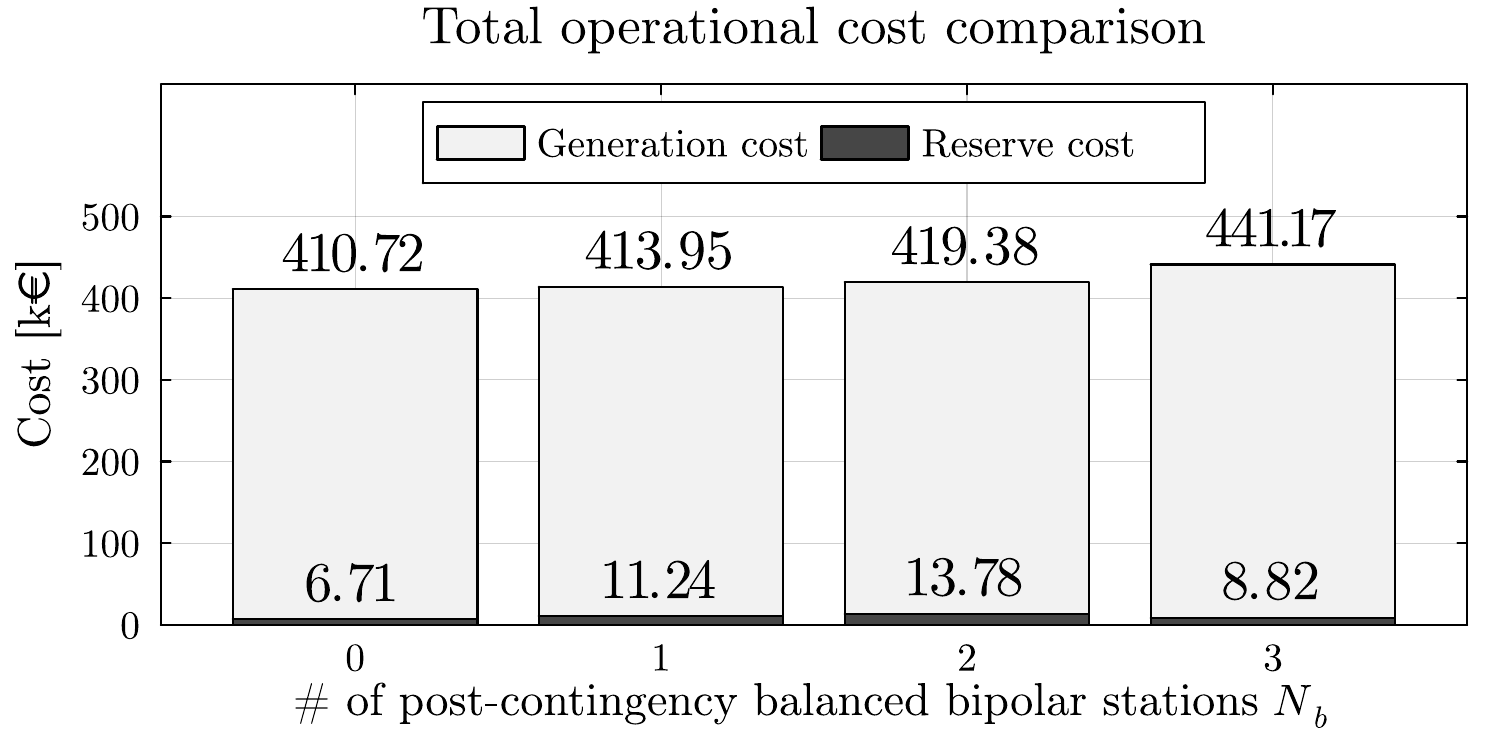}
    \caption{Total operational cost comparison under different $N_{b}$ from the SCOPF model. The reserve costs are also annotated accordingly.}
    \label{fig:cost_compare_SCOPF}
\end{figure}

The results are shown in Fig.~\ref{fig:cost_compare_SCOPF}, where the total operational cost decreases as $N_b$ reduces, exhibiting the same trend as the OPF counterpart.
In the SCOPF case, allowing more converters to operate asymmetrically as a corrective action increases the post-contingency flexibility.
This results in a lower pre-contingency operational cost.
Comparing the two extreme cases, namely $N_b=0$ and $N_b=3$, shows an operational cost reduction of 7.01\%.

A seemingly counterintuitive result is observed in the reserve cost trend.
The reserve cost increases when moving from $N_b = 0$ to $N_b = 2$ and decreases from $N_b = 2$ to $N_b = 3$ while the total operational cost trend persists.
This occurs because limited post-contingency flexibility (less asymmetrical stations) contributes to limiting the amount of reserve required.

This example demonstrates how to quantify the benefits of enabling asymmetrical operation as a post-contingency corrective action, which can be useful for future work to compare the benefits against the potential operational challenges.
For instance, operating a larger number of asymmetrical converter stations may introduce increased complexity in control system coordination.

\subsection{Analysis on the full asymmetrical operation}
\begin{figure}[htbp!]
\vspace{-5mm}
    \centering
    \includegraphics[width=0.6\linewidth]{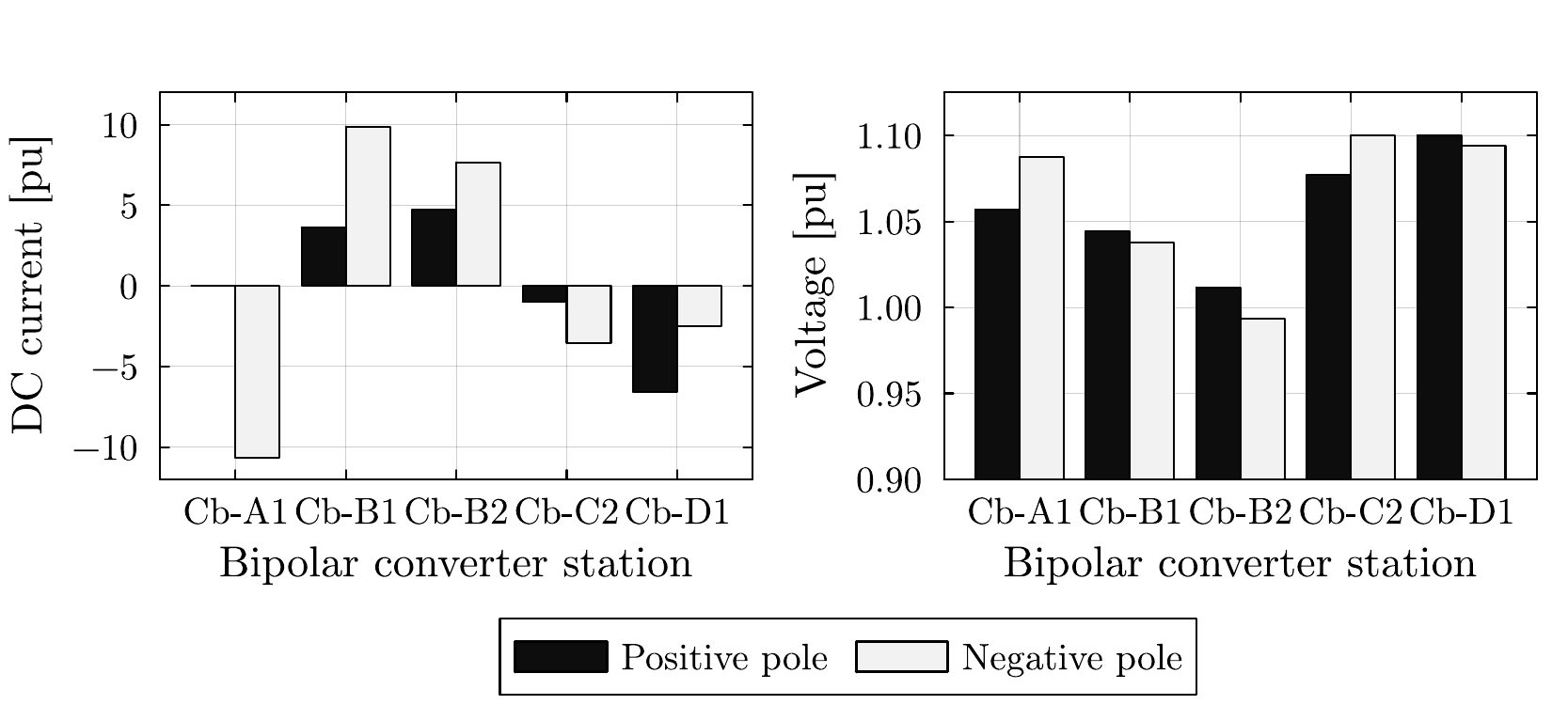}
    \caption{Converter station pole currents and voltages under full asymmetrical operation of the HVDC grid after a converter pole outage of converter station Cb-A1}
    \label{fig:conv_idc_udc_groupedbar}
\end{figure}

We refer to bipolar HVDC grid operation as fully asymmetrical when every converter station connected to the grid operates in an asymmetrical mode.
We take the previous OPF results as an example ($N_b=0$).
The pole currents and voltages of the converter stations are depicted in Fig.~\ref{fig:conv_idc_udc_groupedbar}.
It can be observed that during asymmetrical operation, each pole operates with its own setpoint.
Consequently, compared to symmetrical operation, more careful coordination is required to effectively manage the neutral current flow.

Insufficient coordination may cause excessive voltage drops along the neutral conductor, leading to unsafe DC neutral bus voltage offsets.
The station neutral point, which normally remains close to 0~V, may rise to several kilovolts relative to ground, imposing stress on the insulation and even shock hazards.
This offset may also lead to transformer core saturation due to the resulting common-mode DC flux.

\begin{figure}[htbp!]
    \centering
    \includegraphics[width=0.5\linewidth]{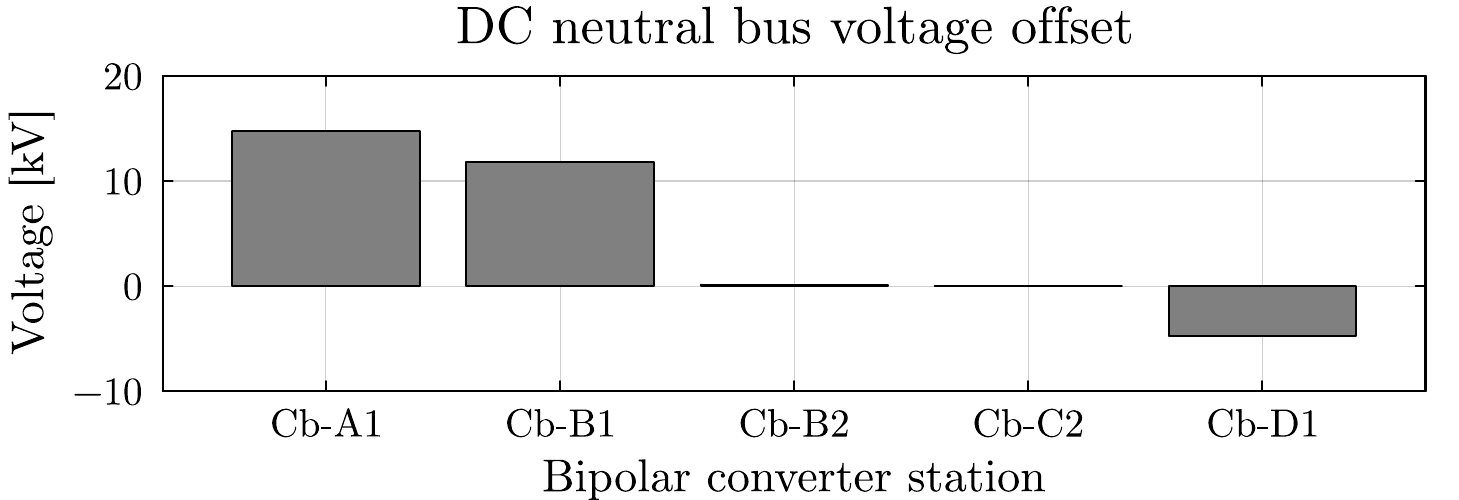}
    \caption{Neutral bus voltage offsets due to currents flowing through the neutral connection.}
    \label{fig:dcneutral_udc}
    \vspace{-2mm}
\end{figure}

These effects are illustrated in Fig.\ref{fig:dcneutral_udc}, which shows the DC neutral bus voltage offsets caused by the neutral currents.
The results indicate that the voltage drop causes the neutral voltage at converter stations CB-A1 and Cb-B1 to exceed 10~kV (400 $\mathrm{kV_{dc}}$ base).
In particular, the offset at converter station Cb-A1 reaches 14.52~kV, corresponding to approximately 3.63\% of the rated pole-to-neutral voltage.
As discussed, such conditions are generally undesirable.
In HVDC grids, the presence of multiple conductors connected to neutral points allows for topological actions that can mitigate this issue and its associated cost impact.

\subsection{DC neutral bus voltage offset limit: Analysis and mitigation via neutral line switching (NLS)}

In this section, we impose DC neutral bus voltage offset limits of 0.02 pu (8 kV) and 0.01 pu (4 kV) to assess their impact on the total operational costs, since this constraint adds an extra limitation to the converter station power transfer, similar to the formulation in~\cite{jatHybridACDC2024}.
To mitigate this, additional flexibility is introduced by reconfiguring the neutral line connections.
This configuration effectively decouples the neutral points and eliminates voltage interdependencies among them.
Again, we use the previous post-contingency OPF results as an example ($N_b=0$).
The reconfiguration is added to the OPF problem by modeling the line status $\gamma$ as a binary decision variable.
The resulting performance metrics are presented in Table~\ref{tab:NLS}.

\def\arraystretch{1.2}
\begin{table}[htbp!]
    \centering
    \caption{Objective value comparison without (base) and with neutral line switching (NLS).}
    \begin{tabular}{lccc}
    \toprule
       \multirow{2}{*}{\shortstack{DC neutral voltage \\ offset limit}} &  \multicolumn{2}{c}{Objective value in \officialeuro} & \multirow{2}{*}{\shortstack{Neutral line\\ disconnected}} \\
        &  Base  &  NLS \\
       \midrule
       Unrestricted & 415,905 & - & -\\
        8 kV  &  422,256  & 420,630 & LD-7 \& LD-9 \\
        4 kV  & 429,400 & 428,152 & LD-7 \& LD-9 \\
       \bottomrule
    \end{tabular}
    \label{tab:NLS}
\end{table}

As shown in the table, the objective value increases when tighter voltage limits are imposed at the neutral points. 
Comparing the unrestricted and 4~kV limit cases highlights this effect, with the operational cost increasing by about 3.2\% to 429,400~\officialeuro.
When reconfiguration through NLS is enabled, this increase is limited to 428,152~\officialeuro.
These results suggest that incorporating NLS into the post-contingency state can provide additional flexibility to maintain acceptable neutral voltage levels while limiting cost increases.

\section{Conclusion}
\label{sec:conclusion}
This paper presents a novel modeling approach and analysis of post-contingency operation in meshed HVDC grids.
The model identifies cost-effective operational strategies by introducing decision variables that determine which converter stations operate asymmetrically following a converter pole outage.
In addition, DC neutral line switching (NLS) is incorporated as a topological action to mitigate DC neutral voltage offsets, demonstrating its role in managing the cost impact as a result of imposing the DC neutral voltage offset limits.
The modeling introduces a novel DC-grid counterpart of the Sparse Tableau Formulation (STF), enabling practical representation of DC grid node-breaker configurations.
This provides a foundation for future work to extend the case study to include HVDC substation reconfiguration strategies using the STF approach, which can be combined with converter mode selection presented in \cite{damanikOptimalConverterControl2024}.
Additionally, the NLS strategy can be enhanced by implementing converter station grounding reconfiguration.

\bibliographystyle{IEEEtran}
\bibliography{references}

\end{document}